\documentclass[
    ,final            
,numberedheadings 
  ]
  {aipproc}

\layoutstyle{8x11single}

\usepackage{subfig}

\newcommand{\Msun}{\ensuremath{M_{\odot}}\,}

\begin{document}

\title{Understanding the X-ray Luminosity Function\\of High Mass X-ray Binaries}


\author{Harshal Bhadkamkar} 
{address={Department of Astronomy and Astrophysics, Tata Institute of Fundamental Research}}

\author{Pranab Ghosh} 
{address={Department of Astronomy and Astrophysics, Tata Institute of Fundamental Research}}


\begin{abstract}
High mass X-ray binary luminosity function (XLF) is an important tool for studying binary evolution 
processes and also the mass loss and consequent evolution in massive stars. We 
calculate the XLF for neutron star binaries using standard scenario for formation and evolution of 
these systems. A one to one relation between primordial binary parameters and the HMXB parameters is 
established. The probability density function is then transformed using the standard Jacobian formalism. 
It is shown that the model successfully explains some basic properties of the observed XLF.
\end{abstract}

\maketitle


%

\section{Primordial Binary Distribution}
\label{sec:pdftrans}

According to the standard formation and evolution scenario, the progenitor system of a high mass X-ray
binary  (HMXB) starts as a 
binary with two massive main sequence stars. The primary is massive enough to form a neutron star at the end of its 
life. Such a binary can be characterized by the masses of primary ($M_p$) and secondary ($M_s$) 
and the initial orbital separation ($a_0$). The orbit is assumed to be circular. These three parameters 
are considered independent of each other. Hence the net probability density function (PDF) of
the primordial binary system is the product of individual PDFs.
The stellar masses are assumed to follow the power-law IMF given by Salpeter (\cite{salpeter}). The same 
power-law index of IMF is taken for both masses with appropriate limits. The orbital separation is assumed to 
follow a loguniform distribution. Hence the net PDF for primordial binaries is,

\begin{equation}
 f_{primo} = \frac{1}{N}\;\frac{(M_p\,M_s)^{-\alpha}}{a_0}
\end{equation}
 
where $\alpha$ is the power-law index of IMF and $N$ is the normalization factor.

\begin{figure}[ht]
  \centering
  \includegraphics[width=0.38\textwidth]{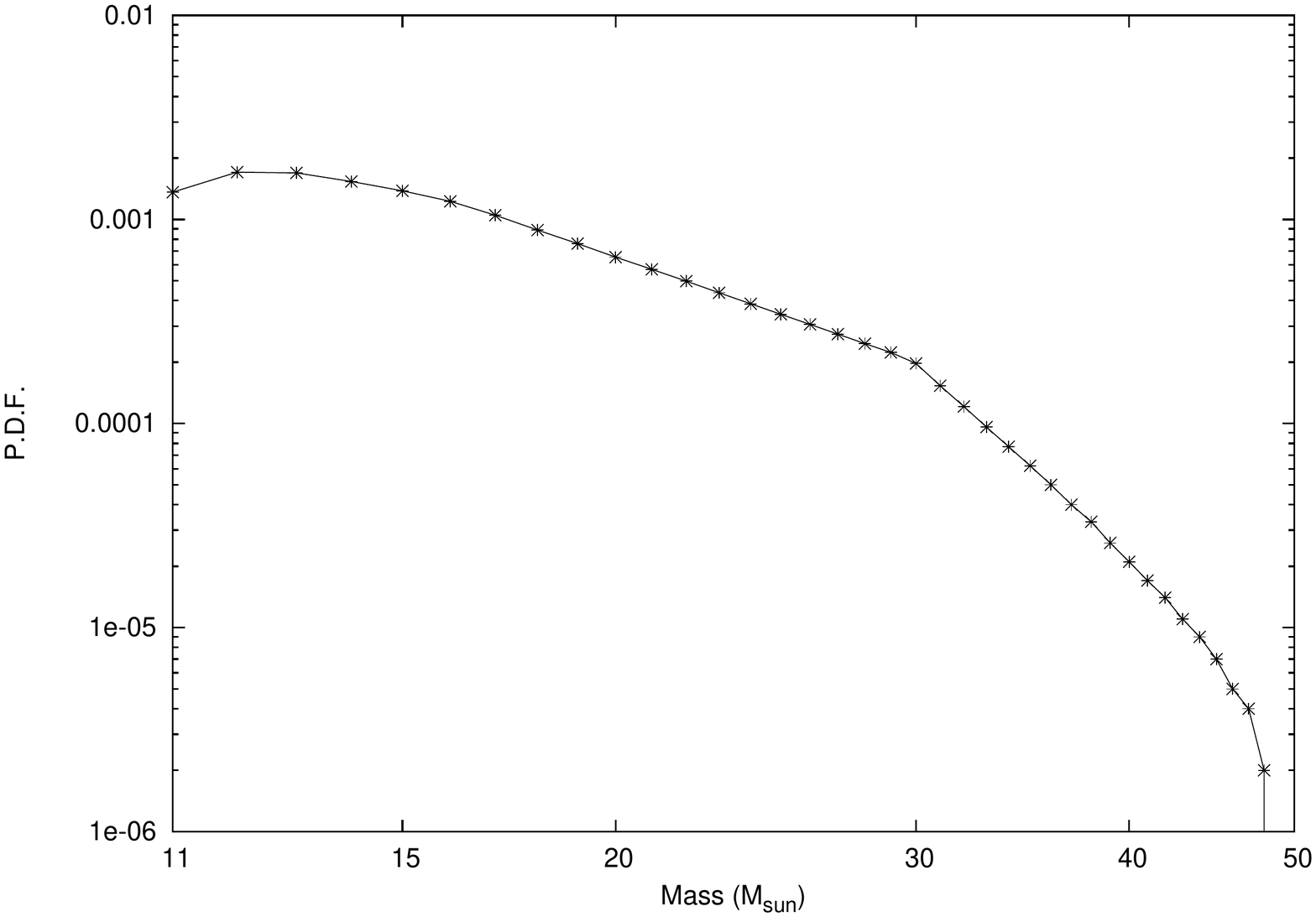}
  \hspace{-30pt}
  \includegraphics[width=0.38\textwidth]{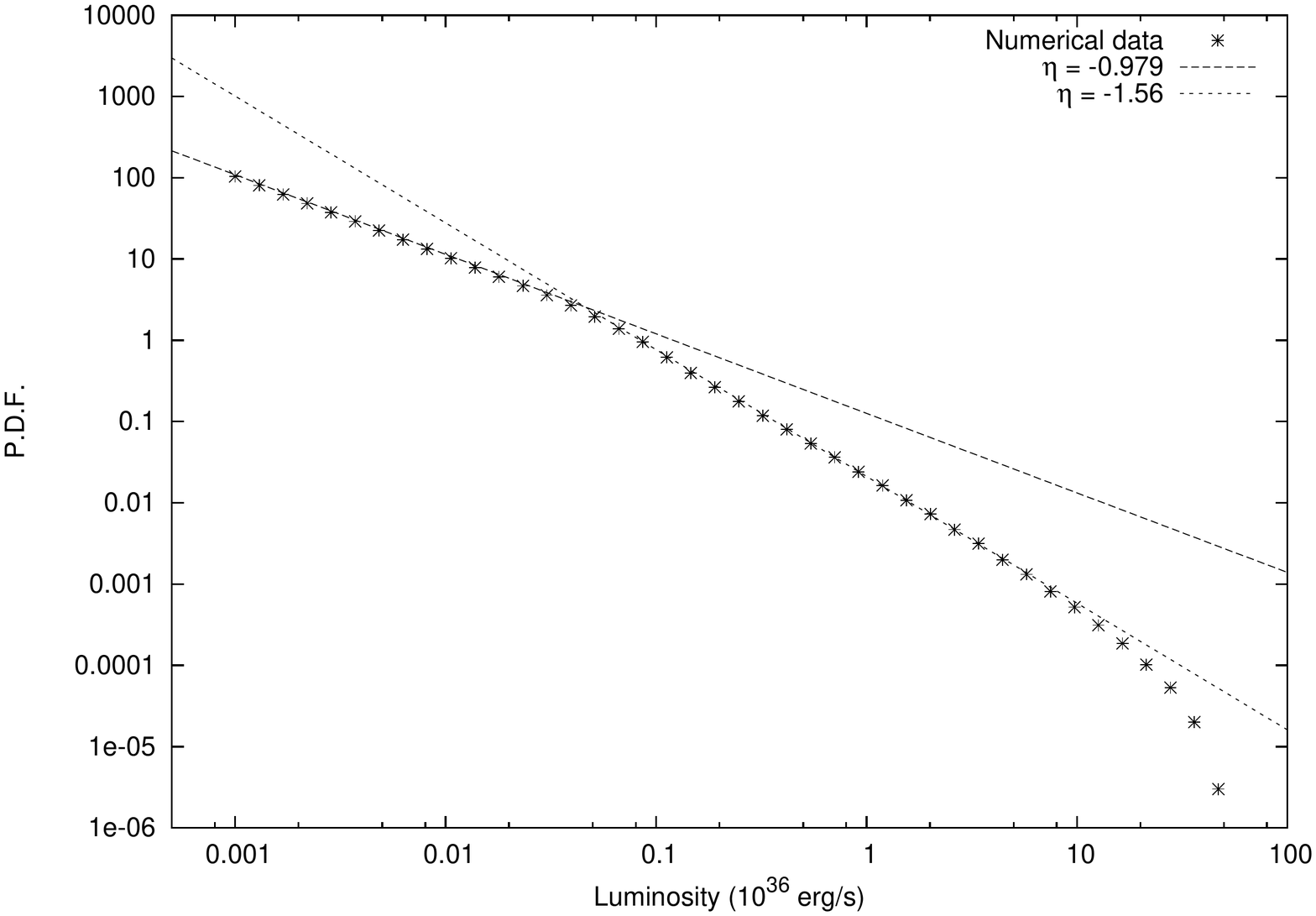}
  \hspace{-30pt}
  \includegraphics[width=0.38\textwidth]{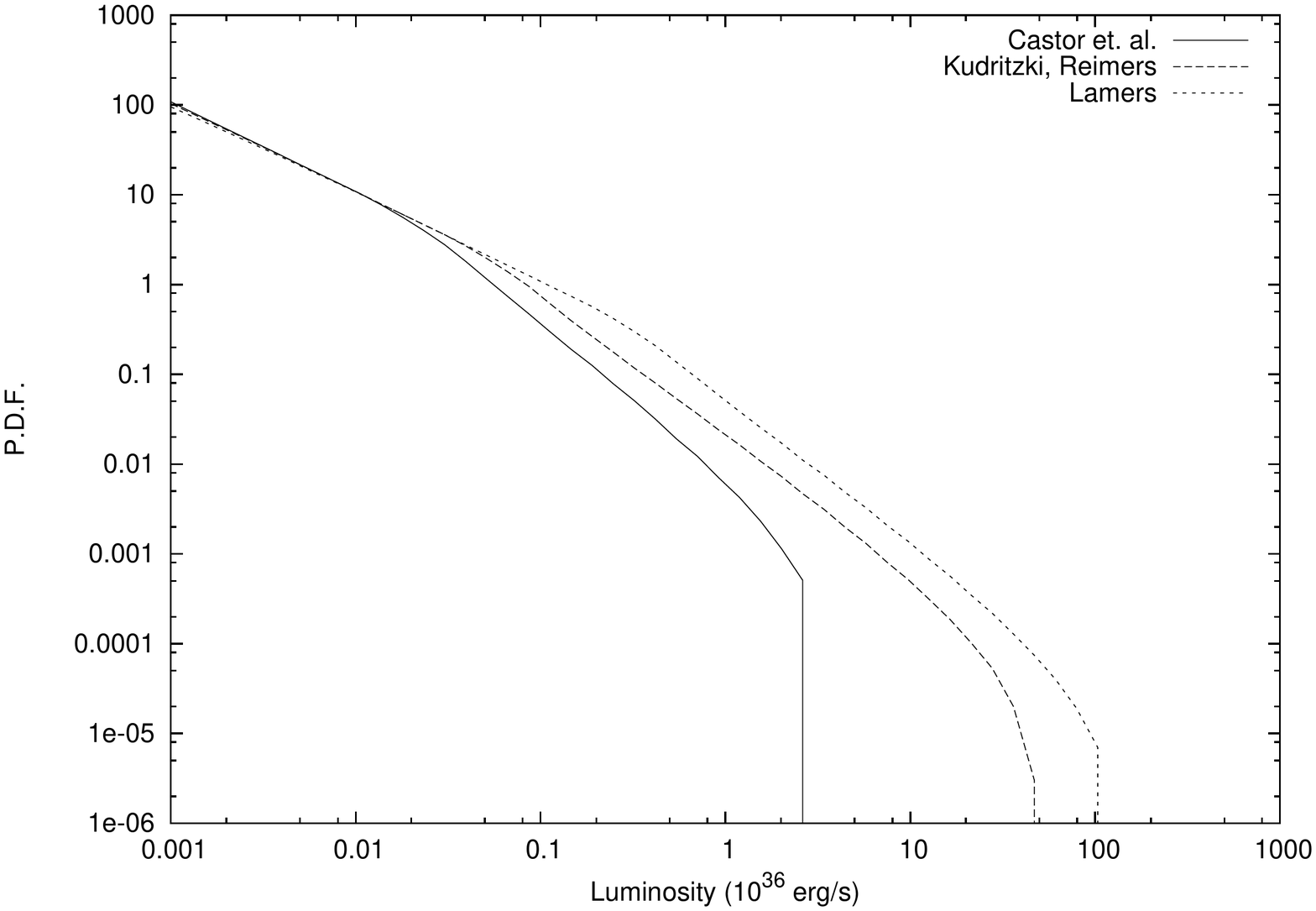}
  \label{fig:figure}
  \caption{(a) Companion Mass PDF, (b) Model Luminosity Function, (c) Effect of wind mass loss model on XLF}
\end{figure}

At the end of the main sequence life of the primary, it expands and fills its Roche lobe. We assume that mass 
transfer in this phase is conservative and the entire envelope of the primary is transfered to the secondary \cite{ghosh}. 
This assumption, along with the relation between the core mass and the total mass of the star (obtained from 
stellar evolution codes), is used to write transformation relations between pre and post mass transfer 
parameters of the system. These relations are then used for transformation of the PDF using Jacobian 
formalism. The core of the primary evolves further and explodes as a supernova. We assume that the mass of 
the neutron star is 1.4 \Msun and rest of the core mass is lost. The post-SN system is characterized by the eccentricity 
($e$), along with the companion mass ($M_c$) and the semimajor axis ($a$). The eccentricity is found to be 
insignificant for HMXBs with a mean value $\sim 0.08$.
Distribution of $M_c$ is interesting being a  broken power-law, as shown in fig. 
\ref{fig:figure} (a). The break occurs at 30 \Msun  which is the upper limit for $M_p$. The power-law index 
of PDF is close to IMF index below 30 \Msun. The PDF falls sharply (with index $\sim 8$) above this value. 

HMXBs are typically wind-fed systems, the companion being near the end of its main sequence life. We use models given 
by various authors for wind mass loss from massive stars and a simplified prescription for the capture 
fraction to calculate the accretion rate (\cite{castor}, \cite{kudritzki}, \cite{lamers}). 
The second parameter for the transformation can be $M_c$ or the 
orbital period $P_b$. The two choices lead to different insights about the HMXB distribution. To 
obtain the XLF we integrate over this second parameter and apply a simple linear transformation 
from accretion rate to luminosity. We refer to our paper for the detailed mathematical treatment. (\cite{harshal})

\section{Properties of the X-Ray Luminosity Function}
\label{sec:xlfprop}

The model XLF has been calculated numerically over a wide luminosity range. It shows a broken power-law with 
a crossover luminosity $L_{cr} \approx 5 \times 10^{34}$ erg/s. The power-law index in bright regime ($L>L_{cr}$)
is $\sim 1.6$ ,
which is in good agreement with observations. For $L<L_{cr}$ the XLF shows a loguniform behaviour. A strong 
cutoff is seen above $10^{37}$ erg/s, which is well within Eddington luminosity for a neutron star. Fig. 
\ref{fig:figure}(b) shows the XLF obtained numerically, the broken power-law so obtained being,

\begin{equation}
\frac{dP}{dL} \propto L^{\eta} \:\mbox{with}\:
\eta = \left\{\begin{array}{ll}
-1.56 & L_{cr}<L_{36}<10 \\
-0.979 & 10^{-3}<L_{36}<L_{cr}
\end{array} \right.
\label{eq:lumpdf}
\end{equation}

The parameters affecting this model XLF are (a) the IMF power-law index, (b) the model chosen for wind 
mass loss rate, (c) the wind velocity and (d) the metallicity of the companion. We discuss here the 
effect of only the wind mass loss model on the
XLF as shown in fig. \ref{fig:figure}(c). We can see that all models overlap for $L<L_{cr}$. This is 
explained by the contribution from the entire range of $M_c$ and hence implies a loguniform nature for 
the XLF. The Castor et. al. model shows a very rapid fall at high $L$, and hence is inapplicable. The other 
two models have similar power-law indices. Thus measurements of $L_{cr}$ can be used to distinguish between
various models.

\section{Conclusion}

We studied the XLF of neutron star HMXBs, obtained by considering the standard model for the formation and 
evolution of HMXBs using PDF transformation formalism. The numerically obtained XLF matches 
well with the observed XLF in bright regime  ($L>L_{cr}$) which has been extensively studied in last decade for nearby
galaxies (\cite{ggs03}, \cite{ggs04}, \cite{ggs05}). The XLF flattens into loguniform shape in faint 
regime ($L<L{cr}$). This 
change has been observed in some recent studies of SMC (\cite{gsh05}). The dependence of XLF power-law index 
on wind mass loss models was also studied. We have excluded from our studies extremely close binaries in which atmospheric 
Roche lobe overflow (ARLOF) is possible. This may extend the cutoff line close to the Eddington limit. 
We emphasize that, in spite of these simplifying assumptions, the calculated XLF matches the observed one 
over most of the luminosity range. 

We plan to include ARLOF systems and black hole binaries into our scheme in future. The observed 
XLF does not appear to show any break or change at the Eddington luminosity for neutron stars. 
This may indicate that the XLF is roughly the same for neutron star and 
black hole binaries, in spite of the differences in detailed accretion mechanism.




\bibliographystyle{aipproc}   




\end{document}